\documentclass{aa}
\usepackage{graphicx}
\usepackage{natbib}
\usepackage{amsmath}
\begin{document}
\authorrunning{Jurcsik et al.}
\title{The Blazhko behaviour of \object{RR Geminorum} I}
\subtitle{CCD photometric results in 2004}

\author{J.~Jurcsik\inst{1}, \'A.~S\'odor\inst{2}, M.~V\'aradi\inst{1}\inst{3},
B.~Szeidl\inst{1}, A.~Washuettl\inst{4}, M.~Weber\inst{4}, I.~D\'ek\'any\inst{2},
Zs.~Hurta\inst{2}, B.~Lakatos\inst{2}, K.~Posztob\'anyi\inst{2}, A.~Szing\inst{2}, \and
K.~Vida\inst{2} 
}

   \institute{Konkoly Observatory of the Hungarian Academy of Sciences. P.O.~Box~67,
H-1525 Budapest, Hungary\\
              \email{jurcsik@konkoly.hu}
   \and E\"otv\"os Lor\'and University, Department of Astronomy, P.O.~Box~32, H-1518
Budapest, Hungary
   \and Department of Experimental Physics and Astronomical Observatory, University of Szeged,
H-6720 Szeged, D\'om t\'er~9, Hungary
   \and Astrophysikalisches Institut Potsdam, An der Sternwarte~16, 14482
Potsdam, Germany }

   \date{Received ; accepted }

\abstract{Extended CCD monitoring of \object{RR Gem} revealed that it is a Blazhko
type RRab star with the shortest Blazhko period ($7\fd23$) and smallest
modulation amplitude
($\Delta M_{max}<0.1$ mag) currently known. The short period of the modulation cycle enabled
us to obtain complete phase coverage of the pulsation at each phase of
the modulation. This is the first multicolour observation of a Blazhko star which is
extended enough to define accurate mean magnitudes and colours of the variable
at different Blazhko phases. Small, but real, changes in the intensity mean colours at
different Blazhko phases have been detected. The Fourier analysis of the light
curves shows that, in spite of the mmag and smaller order of the amplitudes,
the triplet structure is noticeable up to about the 14th harmonic. The modulation is
concentrated to a very narrow, 0.2 phase range of the pulsation, centred on the
supposed onset of the H emission during rising light. These observational
results raise further complications for theoretical explanation of the long known 
but poorly understood Blazhko phenomenon.
  
   \keywords{Stars: individual: \object{RR~Gem} -- 
            Stars: variables: RR~Lyr --
            Stars: oscillations --
            Stars: horizontal-branch --
            Techniques: photometric --
              }
 }  

   \maketitle


\section{Introduction}

Most RR~Lyrae stars repeat their light curves with remarkable regularity. About
30\% of the known galactic RR~Lyrae stars, however, display cyclic modulation in
the shape and amplitude of the light curve over tens to thousands of pulsational
cycles
\citep{smith}. The phenomenon is often called the Blazhko effect since \citet{blazhko}
was the first to recognize that the light maximum of an RR~Lyrae star (namely
\object{RW~Dra})
showed phase modulation on a long time-scale (around 42 days).

In several well-studied RR~Lyrae stars, the amplitude and the period of the Blazhko
effect has been observed to be variable on a time-scale of years
\citep{szeidl,smith99,lacluyze}. In
this respect \object{RR~Geminorum} provides a very interesting case.
\citet{detre70} remarked that his unpublished photographic observations from the
1930s demonstrated the Blazhko effect in \object{RR~Gem}, but (in his
words) it `ceased to exist since about 1940.'
 That the  Blazhko effect might be temporary in nature may be crucial
in understanding the physical background of the phenomenon.
Therefore we decided to begin an
intensive photometric investigation of the star.

\object{RR~Gem} ($\alpha=07^h 21^m 33\fs53, \delta=+30\degr 52\arcmin 59\farcs5$, 
J2000) was discovered by L. P. Ceraski \citep{ceraski} on photographic
plates taken by Blazhko. The first visual observations of the star was made by
\citet{graff} and he was able to determine the pulsation period (0\fd397). Since that
time a great number of visual observations have been
published, but there are only a few published photoelectric observations
available (e.g., \citealp{fitch}; \citealp{liu89}).

The absolute parameters of \object{RR~Gem} were determined by \citet{liu90} from 
a Baade-Wesselink analysis. It belongs to the metal rich ([Fe/H]$=-0.14$~dex
as compiled in \citet{jk96}) disk population group of RR~Lyrae stars
\citep{layden}.
 
Most of the papers dealing with Blazhko stars list the possible theoretical 
explanations of the phenomena and also the different observational
arguments (see e.g. \citealp{lacluyze,macho,chadid}). 
Neither of the two groups of possible models

a) the magnetic oblique rotator (\citealp{shiba,cousens}) 

b) resonant coupling of nonradial mode(s) to the dominant radial mode \citep{nowak1} 

\noindent{can fully explain the large variety of
observational results of Blazhko behaviour. It is important to
note, however, that both of the models connect the Blazhko period to the rotation of
the star.}

In this first part of a series of two papers on \object{RR~Gem}, we examine the
photometric behaviour of the star exclusively using recent CCD observations.
A reexamination of all the previous observations in order to reveal the long term
behaviour of the modulation will be the subject of the second paper. 

\begin{figure}[t]
   \centering
   \includegraphics[width=8.8cm]{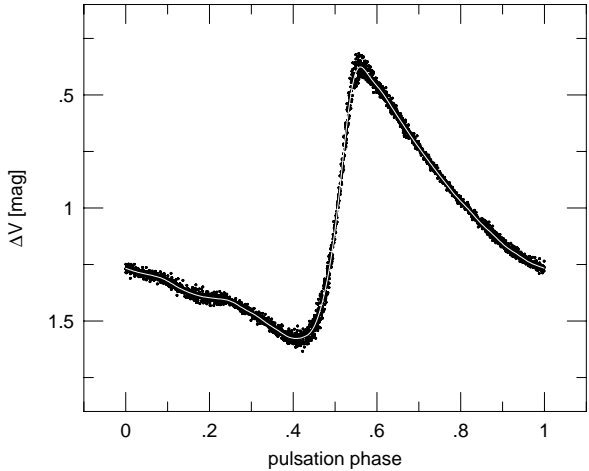}
   \caption{Folded $V$ light curve of \object{RR~Gem}. The scatter of the light
curve is caused by Blazhko modulation and not by observational inaccuracies. 
The amplitude of the modulation is very small, it does not reach 0.1~mag even in
maximum light, and the shape of the modulation points to a very regular behaviour.  
In this, and in the other figures, phase 0.5 is set to the middle of the ascending
branch defined as the phase where the $V$ flux is equal to its time averaged
value.}
              \label{rrgfold}
\end{figure}

\section{The data}

The observations were obtained with the automated 60cm telescope of Konkoly
Observatory, Sv\'abhegy, Budapest equipped with a Wright 750x1100 CCD 
(parameters and calibration are given in \citet{bakos}) using $BV(RI)_c$ filters.
The field of view was 17'x24'.
Data reduction was performed using standard {\sc 
IRAF}\footnotemark[1]\footnotetext[1]{{\sc IRAF} 
is distributed by the National Optical Astronomy Observatories,
    which are operated by the Association of Universities for Research
    in Astronomy, Inc., under cooperative agreement with the National
    Science Foundation.} packages.
More than 3000 frames were obtained in each passband on 56 nights
between 14~January and 4~May in 2004 (JD~$2\,453\,019-2\,453\,130$).
The data were transformed to the
standard system by using the $U, B, V, R, I$ magnitudes given by \citet{liu89} for
\object{SAO~060004} and \object{SAO~060008}. Relative aperture photometry to
\object{BD~$+31\degr1549$} was calculated, and \object{SAO~060004} was used as
check star to control the stability of the results. The magnitude differences
between our comparison and check stars remained
constant with 0.0081, 0.0089, 0.0095, and 0.0093 mag $r.m.s.$ scatter in the $B, 
V, R, I$ colours, respectively.





The magnitudes and colours we derived for \object{BD~$+31\degr1549$}
($V=10.277$~mag, $B-V=0.366$~mag, $V-R=0.233$~mag, and $V-I=0.448$~mag) are in
excellent agreement with the $V=10.273$~mag and $B-V=0.370$~mag values given
by
\citet{sturch}. Photometric data are available electronically at the CDS. 
In the electronic tables
(http://cdsweb.u-strasbg.fr/cats/J.A+A.htx.....)
Column 1 lists the HJD of the observation, and Column 2 gives the
differential magnitude of \object{RR~Gem} with respect to \object{BD~$+31\degr1549$}
for the $B,V, R, I$ colours.

\section{Photometric results}

\begin{table*}[ttt!!!]
   \caption{Amplitudes and phases of the pulsation and modulation frequency
components of \object{RR~Gem}.}
         \label{four}
         \begin{tabular}{rrr@{\hspace{10pt}}rrrrrrr}
            \hline\hline
            \noalign{\smallskip}
&& \multicolumn{2}{c}{$B$} & \multicolumn{2}{c}{$V$} & \multicolumn{2}{c}{$R$}
& \multicolumn{2}{c}{$I$}\\
            \noalign{\smallskip}
            \hline
            \noalign{\smallskip}
\multicolumn{2}{c}{$A_0$ / rms [mag]} & 10.822 & 0.013 & 11.418 & 0.011 &
11.152 & 0.013 & 10.906 & 0.012 \\
	    \noalign{\smallskip}
            \hline
            \noalign{\smallskip}

\multicolumn{2}{c}{$f$} & \multicolumn{1}{c}{A} & \multicolumn{1}{c}{$\phi$} &
\multicolumn{1}{c}{A} & \multicolumn{1}{c}{$\phi$} & \multicolumn{1}{c}{A} &
\multicolumn{1}{c}{$\phi$} & \multicolumn{1}{c}{A} & \multicolumn{1}{c}{$\phi$}

\\

\multicolumn{2}{c}{[c/d]} & \multicolumn{1}{c}{[mag]} & \multicolumn{1}{c}{[deg]} &
\multicolumn{1}{c}{[mag]} & \multicolumn{1}{c}{[deg]} & \multicolumn{1}{c}{[mag]} & 
\multicolumn{1}{c}{[deg]} & \multicolumn{1}{c}{[mag]} & \multicolumn{1}{c}{[deg]} \\

	    \noalign{\smallskip}
            \hline
            \noalign{\smallskip}
$f_0$ & 2.517063 & 0.5527 & 80.0 & 0.4009 & 77.1 & 0.3080 & 72.9 & 0.2376 & 66.7 \\
2$f_0$ & 5.034126 & 0.3111 & 296.1 & 0.2303 & 295.8 & 0.1799 & 294.7 & 0.1395 & 293.2 \\
3$f_0$ & 7.551189 & 0.1751 & 169.7 & 0.1327 & 169.9 & 0.1055 & 169.6 & 0.0828 & 169.5 \\
4$f_0$ & 10.068252 & 0.1184 & 46.2 & 0.0900 & 46.1 & 0.0707 & 45.6 & 0.0557 & 45.9 \\
5$f_0$ & 12.585315 & 0.0663 & 278.0 & 0.0509 & 277.7 & 0.0409 & 276.4 & 0.0322 & 277.8 \\
6$f_0$ & 15.102378 & 0.0494 & 142.9 & 0.0385 & 142.9 & 0.0304 & 143.1 & 0.0241 & 142.2 \\
7$f_0$ & 17.619441 & 0.0341 & 18.3 & 0.0265 & 18.9 & 0.0212 & 18.7 & 0.0165 & 18.3 \\
8$f_0$ & 20.136504 & 0.0217 & 239.5 & 0.0170 & 241.4 & 0.0133 & 241.5 & 0.0099 & 241.8 \\
9$f_0$ & 22.653567 & 0.0174 & 105.1 & 0.0134 & 108.9 & 0.0109 & 109.5 & 0.0086 & 109.2 \\
10$f_0$ & 25.170630 & 0.0125 & 333.2 & 0.0098 & 335.4 & 0.0071 & 338.7 & 0.0057 & 330.0 \\
11$f_0$ & 27.687693 & 0.0091 & 196.2 & 0.0074 & 198.9 & 0.0054 & 195.5 & 0.0034 & 192.9 \\
12$f_0$ & 30.204756 & 0.0072 & 57.2 & 0.0048 & 58.34 & 0.0044 & 58.6 & 0.0030 & 61.0 \\ 
13$f_0$ & 32.721819 & 0.0055 & 282.0 & 0.0042 & 280.3 & 0.0036 & 294.2 & 0.0021 & 285.9 \\
14$f_0$ & 35.238882 & 0.0044 & 145.8 & 0.0031 & 148.2 & 0.0026 & 141.3 & 0.0015 & 149.1 \\
15$f_0$ & 37.755945 & 0.0039 & 3.9 & 0.0027 & 1.2 & 0.0024 & 2.7 & 0.0013 & 359.1 \\
$f_0-f_m$ & 2.378743 & 0.0093 & 185.6 & 0.0065 & 186.6 & 0.0059 & 196.2 & 0.0044 & 192.6 \\
$f_0+f_m$ & 2.655383 & 0.0092 & 349.2 & 0.0064 & 347.3 & 0.0058 & 350.9 &0.0039 & 352.4 \\
2$f_0-f_m$ & 4.895806 & 0.0092 & 29.3 & 0.0073 & 38.8 & 0.0052 & 32.6 & 0.0044 & 42.7 \\
2$f_0+f_m$ & 5.172446 & 0.0086 & 175.3 & 0.0060 & 179.1 & 0.0043 & 181.2 & 0.0039 & 186.5 \\
3$f_0-f_m$ & 7.412869 & 0.0078 & 248.6 & 0.0061 & 248.4 & 0.0052 & 262.8 & 0.0038 & 271.1 \\
3$f_0+f_m$ & 7.689509 & 0.0071 & 43.9 & 0.0052 & 43.7 & 0.0046 & 45.8 & 0.0036 & 54.4 \\
4$f_0-f_m$ & 9.929932 & 0.0070 & 155.3 & 0.0054 & 156.3 & 0.0040 & 149.5 & 0.0032 & 161.5 \\
4$f_0+f_m$ & 10.206572 & 0.0067 & 306.8 & 0.0048 & 302.8 & 0.0043 & 304.1 & 0.0035 & 303.5 \\
5$f_0-f_m$ & 12.446995 & 0.0056 & 18.4 & 0.0043 & 14.8 & 0.0034 & 16.3 & 0.0028 & 16.4 \\
5$f_0+f_m$ & 12.723635 & 0.0054 & 190.0 & 0.0042 & 185.6 & 0.0033 & 188.6 & 0.0030 & 183.6 \\
6$f_0-f_m$ & 14.964058 & 0.0060 & 253.1 & 0.0043 & 250.4 & 0.0034 & 255.9 & 0.0027 & 243.3 \\
6$f_0+f_m$ & 15.240698 & 0.0035 & 41.3 & 0.0037 & 40.1 & 0.0024 & 38.1 & 0.0018 & 56.0 \\
7$f_0-f_m$ & 17.481121 & 0.0049 & 135.6 & 0.0036 & 136.1 & 0.0029 & 128.2 & 0.0027 & 126.9 \\
7$f_0+f_m$ & 17.757761 & 0.0048 & 280.5 & 0.0040 & 279.3 & 0.0037 & 282.5 & 0.0021 & 279.7 \\
8$f_0-f_m$ & 19.998184 & 0.0033 & 358.2 & 0.0026 & 1.2 & 0.0020 & 4.0 & 0.0026 & 356.5 \\
8$f_0+f_m$ & 20.274824 & 0.0033 & 169.6 & 0.0027 & 168.2 & 0.0018 & 172.1 & 0.0018 & 174.1 \\
9$f_0-f_m$ & 22.515247 & 0.0036 & 245.5 & 0.0025 & 240.8 & 0.0031 & 242.6 & 0.0017 & 246.8 \\
9$f_0+f_m$ & 22.791887 & 0.0030 & 30.6 & 0.0020 & 29.8 & 0.0017 & 21.4 & 0.0010 & 34.8 \\
10$f_0-f_m$ & 25.032310 & 0.0025 & 114.8 & 0.0020 & 100.3 & 0.0018 & 130.7 & 0.0014 & 113.8 \\
10$f_0+f_m$ & 25.308950 & 0.0024 & 269.4 & 0.0025 & 263.3 & 0.0012 & 254.8 & 0.0008 & 288.3 \\
11$f_0-f_m$ & 27.549373 & 0.0020 & 332.5 & 0.0017 & 339.3 & 0.0014 & 2.6 & 0.0014 & 325.4 \\
11$f_0+f_m$ & 27.826013 & 0.0018 & 138.7 & 0.0012 & 141.8 & 0.0012 & 152.6 & 0.0009 & 150.7 \\
12$f_0-f_m$ & 30.066436 & 0.0019 & 205.0 & 0.0009 & 205.5 & 0.0014 & 223.3 & 0.0011 & 206.3 \\
12$f_0+f_m$ & 30.343076 & 0.0016 & 356.0 & 0.0008 & 8.6 & 0.0007 & 343.0 &0.0013 & 19.0 \\
13$f_0-f_m$ & 32.583499 & 0.0020 & 115.9 & 0.0015 & 111.0 & 0.0010 & 103.8 & 0.0003 & 147.5 \\
13$f_0+f_m$ & 32.860139 & 0.0019 & 252.5 & 0.0013 & 237.3 & 0.0006 & 223.0 & 0.0010 & 255.9 \\
14$f_0-f_m$ & 35.100562 & 0.0008 & 344.8 & 0.0008 & 330.2 & 0.0009 & 315.7 & 0.0006 & 329.6 \\
14$f_0+f_m$ & 35.377202 & 0.0011 & 155.2 & 0.0009 & 159.8 & 0.0013 & 122.4 & 0.0009 & 101.7 \\
15$f_0-f_m$ & 37.617625 & 0.0005 & 161.0 & 0.0007 & 219.6 & 0.0006 & 178.5 & 0.0006 & 140.6 \\
15$f_0+f_m$ & 37.894265 & 0.0008 & 11.8 & 0.0004 & 354.5 & 0.0005 & 30.4 & 0.0002 & 155.3 \\
$f_m$ & 0.138320 & 0.0058 & 345.9 & 0.0045 & 355.7 & 0.0040 & 345.9 & 0.0042 & 354.1 \\

            \noalign{\smallskip}
            \hline
         \end{tabular}
\end{table*}

\begin{table*}[tttthh!!!!]
   \caption{Side lobe amplitude ratios, phase differences, and their errors.
$R_k=A_{kf_0+f_m}/A_{kf_0-f_m}; \Delta \phi_k=
\phi_{kf_0+f_m}-\phi_{kf_0-f_m}$.}
         \label{rfi}

\begin{tabular}{r@{\hspace{10pt}}
                r@{\hspace{8pt}}r@{\hspace{8pt}}r@{\hspace{8pt}}r@{\hspace{16pt}}
                r@{\hspace{8pt}}r@{\hspace{8pt}}r@{\hspace{8pt}}r@{\hspace{16pt}}
                r@{\hspace{8pt}}r@{\hspace{8pt}}r@{\hspace{8pt}}r@{\hspace{16pt}}
                r@{\hspace{8pt}}r@{\hspace{8pt}}r@{\hspace{8pt}}r}
            \hline\hline
            \noalign{\smallskip}
 & \multicolumn{4}{c}{$B$} & \multicolumn{4}{c}{$V$} & \multicolumn{4}{c}{$R$}
& \multicolumn{4}{c}{$I$} \\
            \noalign{\smallskip}
            \hline
            \noalign{\smallskip}
$k$ & 
\multicolumn{1}{c}{$R_k$} & \multicolumn{1}{c}{$\sigma_{R_k}$} &
\multicolumn{1}{c}{$\Delta\phi_k$} &
\multicolumn{1}{l}{$\sigma_{\Delta\phi_k}$} &
\multicolumn{1}{c}{$R_k$} & \multicolumn{1}{c}{$\sigma_{R_k}$} &
\multicolumn{1}{c}{$\Delta\phi_k$} &
\multicolumn{1}{l}{$\sigma_{\Delta\phi_k}$} &
\multicolumn{1}{c}{$R_k$} & \multicolumn{1}{c}{$\sigma_{R_k}$} &
\multicolumn{1}{c}{$\Delta\phi_k$} &
\multicolumn{1}{l}{$\sigma_{\Delta\phi_k}$} &
\multicolumn{1}{c}{$R_k$} & \multicolumn{1}{c}{$\sigma_{R_k}$} &
\multicolumn{1}{c}{$\Delta\phi_k$} &
\multicolumn{1}{l}{$\sigma_{\Delta\phi_k}$} \\
            \noalign{\smallskip}
            \hline
            \noalign{\smallskip}

1 & 0.989 & 0.045 & 163.6 & 3.0 & 0.970 & 0.063 & 160.7 & 3.7 
  & 0.983 & 0.071 & 154.7 & 4.9 & 0.867 & 0.088 & 159.8 & 6.2  \\
2 & 0.934 & 0.053 & 146.0 & 3.3 & 0.822 & 0.053 & 140.3 & 3.8 
  & 0.827 & 0.100 & 148.6 & 6.3 & 0.886 & 0.091 & 143.8 & 6.5  \\
3 & 0.910 & 0.060 & 155.3 & 3.7 & 0.852 & 0.065 & 155.3 & 4.1
  & 0.885 & 0.103 & 143.0 & 5.7 & 0.947 & 0.109 & 143.3 & 6.8  \\
4 & 0.957 & 0.070 & 151.5 & 3.9 & 0.887 & 0.076 & 146.5 & 4.6
  & 1.075 & 0.128 & 154.6 & 6.6 & 1.063 & 0.137 & 142.0 & 7.4  \\
5 & 0.982 & 0.088 & 171.6 & 5.2 & 0.977 & 0.098 & 170.8 & 5.7
  & 0.971 & 0.144 & 172.3 & 8.8 & 1.034 & 0.176 & 167.2 & 9.0  \\
6 & 0.583 & 0.077 & 148.2 & 6.5 & 0.860 & 0.092 & 149.7 & 5.9
  & 0.706 & 0.144 & 142.2 & 10.3 & 0.667 & 0.134 & 172.7 & 11.8  \\
7 & 0.980 & 0.114 & 144.9 & 6.0 & 1.111 & 0.125 & 143.2 & 6.5
  & 1.276 & 0.224 & 154.3 & 9.1 & 0.778 & 0.141 & 152.8 & 11.5  \\
8 & 1.000 & 0.129 & 171.4 & 8.6 & 1.038 & 0.166 & 167.0 & 9.2
  & 0.900 & 0.269 & 168.1 & 15.5 & 0.692 & 0.140 &177.6 & 12.7  \\
9 & 0.833 & 0.108 & 145.1 & 8.9 & 0.800 & 0.154 & 149.0 & 11.2
  & 0.548 & 0.110 & 138.8 & 13.8 & 0.588 & 0.205 & 148.0 & 22.5  \\
10 & 1.000 & 0.170 & 154.6 & 11.4 & 1.250 & 0.240 & 163.0 & 10.7
   & 0.632 & 0.187 & 124.1 & 20.6 & 0.714 & 0.263 & 174.5 & 24.0  \\

            \noalign{\smallskip}
            \hline
         \end{tabular}
\end{table*}

\begin{figure}[t]
   \centering
   \includegraphics[width=8.8cm]{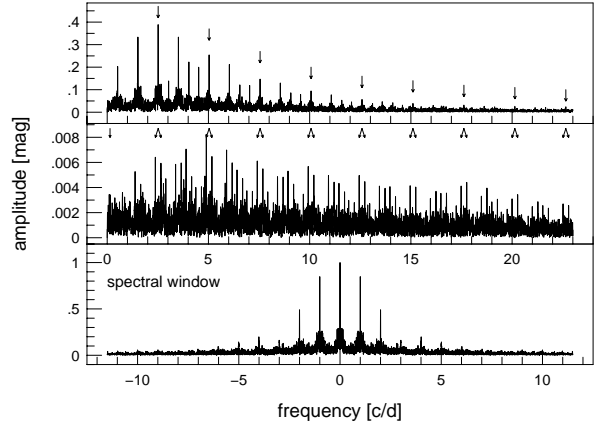}
   \caption{Amplitude spectrum of the $V$ light curve (top panel).
Middle panel shows the residual spectrum after prewhitening with the
pulsation frequency and its harmonics ($kf_0,~k=1...15$).
The location of the modulation frequency ($f_m$) and the symmetrically placed
modulation frequency side lobes ($kf_0 \pm f_m$) are indicated by arrows. The spectral
window is shown in the bottom panel.
}
              \label{rrgsp}
\end{figure}

\begin{figure}[]
   \centering
   \includegraphics[width=8.3cm]{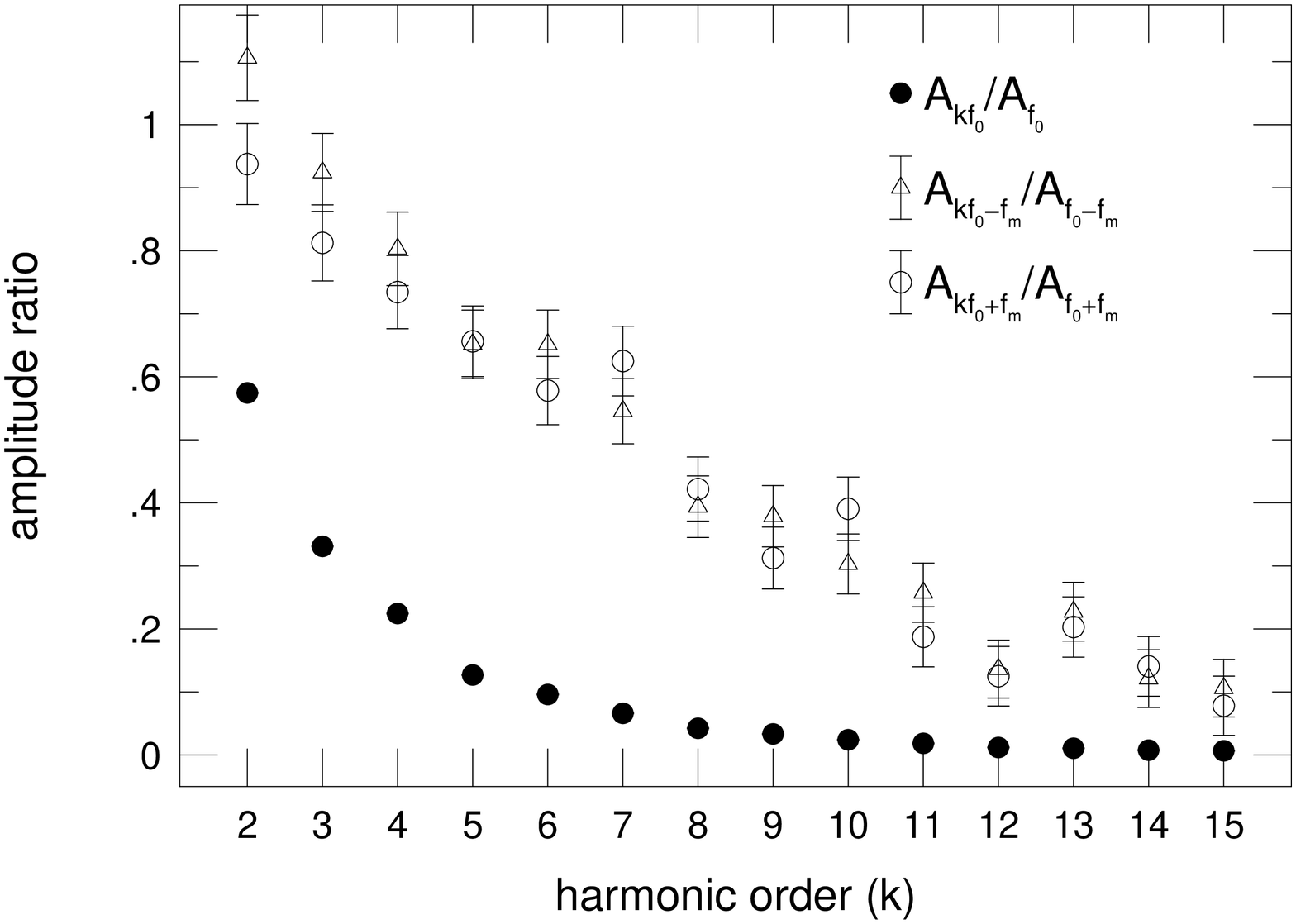}
   \caption{Amplitude ratios of the harmonic components of the dominant radial 
pulsation mode $A_{kf_0}/A_{f_0}$ (solid) compared to the amplitude ratio
behaviour of the modulation components $A_{kf_0-f_m}/A_{f_0-f_m}, 
A_{kf_0+f_m}/A_{f_0+f_m}$ (open symbols). While the amplitudes of the pulsation
decrease exponentially, the amplitude decrease of the modulation is linear. Error bars
for the modulation components' amplitude ratios are indicated, the errors of the
amplitude ratios of the dominant pulsation mode are smaller than the symbols used.}
              \label{amp2}
\end{figure}

The CCD observations revealed that, indeed, \object{RR~Gem} shows Blazhko
modulation, however, with very small, $\sim$~$0.1$ mag modulation amplitude. The
modulation pattern is very regular, with a definite node (fixed point) on the
ascending branch. Fig.~\ref{rrgfold} shows the $V$ band observations superimposed
on the mean light curve. The variation in the heights of maxima indicated a
modulation period of about 7~days.

The Fourier spectra of the $V$ data and the data prewhitened with the pulsation
frequency and its harmonics are shown in Fig.~\ref{rrgsp}. Data analysis was
performed using the different applications of the MUFRAN package \citep{mufran}. 
The Fourier spectrum of the residual shows the triplet structure typical
of amplitude modulation up to about the 14th harmonic component. The side lobe 
frequencies have similar amplitudes and are symmetrically placed around the
radial mode frequency and its harmonics. The amplitudes of the side lobes decrease
much less steeply than the amplitudes of the harmonics of the main pulsation
frequency (see Fig.~\ref{amp2}). This fact may imply a different explanation of
these frequency components than  that of the radial mode where nonlinear effects
invoke harmonic components with exponentially decreasing amplitudes.
The modulation components' amplitudes of \object{AR~Her} (a Blazhko variable with
large amplitude and phase modulation and a 32~d modulation period) show similar
behaviour; from the fourth harmonic component its modulation amplitudes dominate
over the pulsation ones \citep{smith99}.

Table~\ref{four} summarizes the amplitudes and phases of the pulsation and
modulation frequencies in the $B, V, R, I$ passbands calculated from the least squares
solution using sine terms and HJD~$T_0=2\,453\,019.0$ initial epoch value
according to the formula given in \citet[Eq.~2]{kovacs}. In the top of
Table~\ref{four} the mean magnitudes ($A_0$) and the standard deviations  of the
fits are also given.

\begin{figure}[hhhh]
   \centering
   \includegraphics[width=8.8cm]{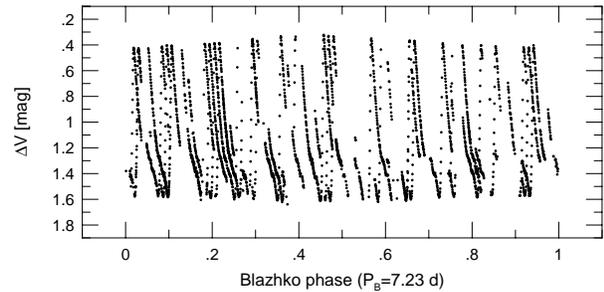}
   \caption{The $V$ light curve of \object{RR~Gem} folded with the $7\fd23$
Blazhko period. The heights of maximum light clearly show a 0.1 mag amplitude
modulation.   
}
              \label{phase}
\end{figure}

This solution has been calculated by using locked values of both the pulsation
and modulation frequency components. The pulsation frequency adopted is the mean
value obtained for the $B, V, R, I$ data. The modulation frequency has been
determined as the frequency which results the smallest $r.m.s.$ scatter of the
residuals of the least squares solutions assuming symmetrical displacement of the
side lobes up to the 9th harmonic component of the radial mode. The period of the
modulation obtained this way is $7\fd23$. In Fig.~\ref{phase} the $V$ light curve
is folded according to this modulation period; sinusoidal variation in the height
of maximum light is evident. The ephemerides of maximum pulsation light and
maximum pulsation amplitude determined from the measurements of the 2004 
observing season are the followings: 

HJD~$T_{max} =  2\,453\,019.1590 + 0.3972884\cdot{\rm {E_{pulsation}}}$

HJD~$T_{Blmax} = 2\,453\,021.30  + 7.23\cdot{\rm {E_{Blazhko}}}$.

The error weighted mean value of the asymmetry parameter   
($Q={{A_+-A_-}\over{A_++A_-}}$) averaged over the first 10 modulation component pairs
in the four colours is $-0.035\pm0.057$, indicating very small asymmetry
with slightly larger amplitudes at the shorter frequency sides. 
For comparison the distribution of this parameter of the MACHO data 
peaks at $+0.3$ \citep{macho}.

\begin{figure*}[ttttttthhhhh!!!!!]
   \centering
   \includegraphics[width=18cm]{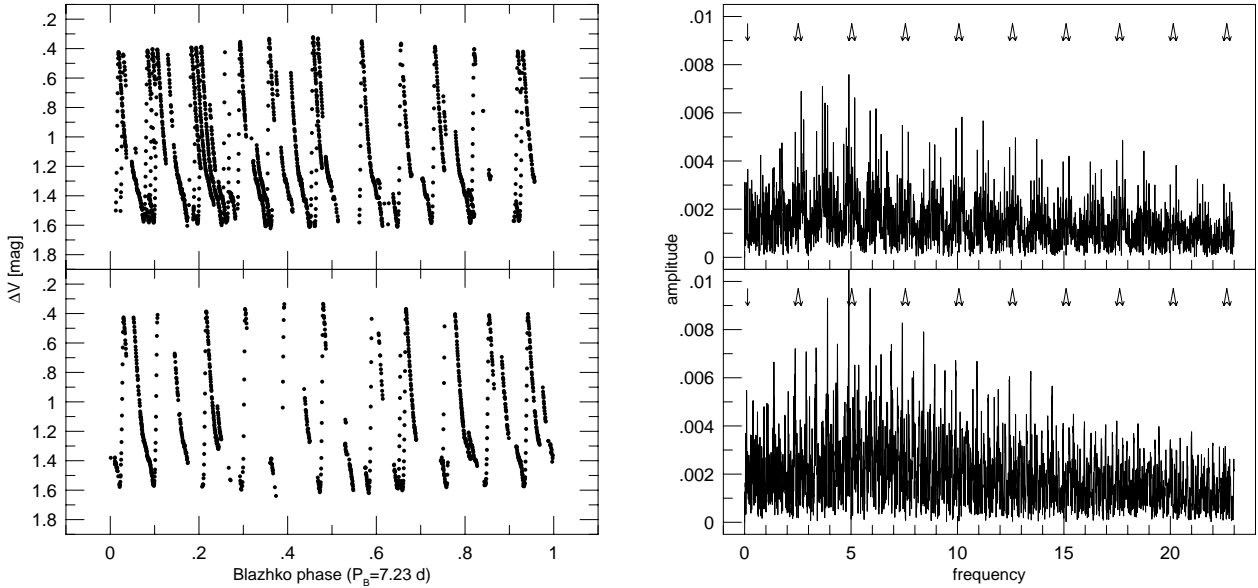}
   \caption{$V$ data folded with the Blazhko period (left panels)
and their residual spectra after prewhitening with the pulsation period and its
harmonics (right panels) for two equal length parts of the dataset.
In the first half of the observations (between JD 2\,453\,019 and 2\,453\,174) the
data coverage is nearly the same as for the whole data set and the residual
spectrum is very similar to that shown in Fig.~\ref{rrgsp} for the complete data
set (top panels). In the second part of the observations (between JD~2\,453\,077
and~2\,453\,130) the data are more sparse, and the residual spectrum has changed
significantly. The near equality of the sidelobe amplitudes has disappeared, instead,
highly asymmetrical structure emerge, with very small if any power at the  higher
frequency sides (bottom panels).
}
              \label{phase2}
\end{figure*}

The currently viable theoretical explanations of the Blazhko phenomenon
predict
approximate equality of the amplitude ratios, and phase differences between
the side lobe frequencies of the different order components in each colour
\citep{kovacs,shiba,smith99,nowak1}.
Although these predictions are based on simplified models it is worth checking
the stability of these components in our data. Table~\ref{rfi} lists the amplitude
ratios and  phase differences and their errors of the side lobe frequency pairs in
the $B, V, R, I$ bandpasses for the first 10 modulation component pairs.
Differences larger than $2\sigma$ can be found in the amplitude ratios while the
phase differences of the first 5 modulation component pairs in the different
colours do not agree within the $3\sigma$ limit. The mean values and the standard
deviations of the phase differences in the $B, V, R, I$ colours of the first five
modulation component pairs are $159.7\pm3.7, 144.7\pm3.5, 149.2\pm7.0,
148.7\pm5.5,$ and $170.5\pm2.3$. The phase differences of the fifth components 
are statistically significantly larger in each colour than the phase differences
of the second pairs. The large scatter of the third and fourth
components  indicates colour to colour changes of the phase differences, too.
Differences between these parameters ($R_k$,
$\Delta \phi_k$) have also been detected in \object{AR~Her} \citep{smith99}.

\subsection{Stability of the results}

One of the theoretically unexplained characteristics of Blazhko stars
is the high asymmetry of the amplitudes of the $\pm$ modulation frequency
components. In contrast, only very small if any asymmetry can be seen in
Fig.~\ref{rrgsp}. We have checked what  happens if data subsets with less complete
phase coverage are used. The results are shown in Fig.~\ref{phase2}.
The 110 day dataset has been divided into two intervals of equal length but
different number of observations ($\sim$~$2000$ and $\sim$~$1000$).
In the first part of the observations both the clear sky statistics were better, and
the observational runs were longer. The residual spectrum of the first dataset is
very similar to that of the complete dataset shown in the middle panel of
Fig.~\ref{rrgsp}. However, the residual spectrum of the second dataset is
significantly different, with enhanced power at the lower frequency side of the
modulation, and with diminishing amplitudes at the higher frequency side. 
The data distribution of this second dataset also contains observations
from each 0.1 phase intervals of the Blazhko period, but for some phase intervals 
no complete light curve can be constructed. This is the typical data distribution
of Blazhko star photometry, thus our result warns that classification
schemes
of the modulation based on the asymmetry of the side lobe amplitudes  might be
seriously biased by the phase coverage of the data and may lead to an
over-interpretation of the results. As the Blazhko period of \object{RR~Gem} is
about an order of magnitude shorter than the typical period of the modulation,
it should be expected that for stars with longer modulation periods
 even  more extended observations than we have on \object{RR~Gem} are needed 
to correctly describe the properties of the modulation. For comparison, in
\citet{macho}, $700-1000$ measurements of the stars were analysed.

The effect of data distribution biases on the side lobe amplitudes, however, has 
to be carefully treated. The fact that the different observations
of \object{RR~Lyrae} (\citealp{szk}; \citealp{smith03}) equally show the longer 
frequency modulation components to have higher amplitudes points to real
differences in the amplitudes of the modulation components.

\section{Variations during the Blazhko cycle}

The full coverage of the pulsation light curve at different phases of the Blazhko
cycle makes it possible to study the changes of the light curves' shapes and
Fourier parameters in detail, and to detect any possible variation in the mean
magnitudes and colours.

\subsection{Light curve changes}

\begin{figure}[bbbb!!!!!]
   \centering
   \includegraphics[width=8.8cm]{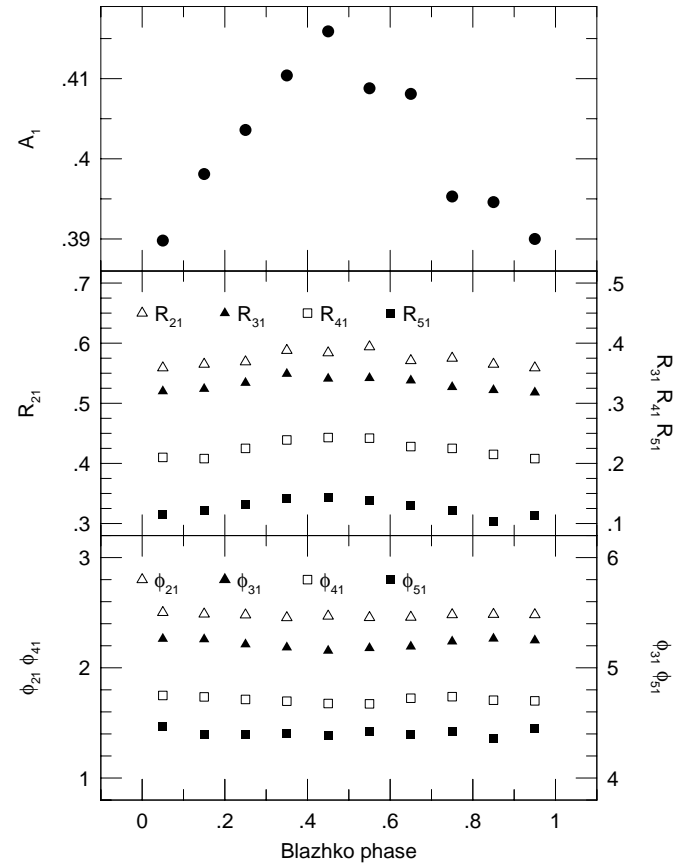}
   \caption{Fourier parameters of the $V$ light curves in the different
phases of the Blazhko cycle. The amplitude ratios vary in phase with
the $A_1$ amplitude, but no changes in the epoch independent phase differences can
be detected. }
              \label{afi}
\end{figure}

We have divided the data into 10 subsets containing observations from different
phases of the Blazhko cycle. The minimum and maximum  number of data in these
subsets are 165 and 479, respectively. The behaviour of the Fourier parameters of
the light curves in the different Blazhko phases is shown in Fig.~\ref{afi}.
The total range of the variations in the $A_1$, $A_{10}$ amplitudes are 0.026,
(between 0.390 and 0.416~mag) and 0.009~mag (between 0.005 and 0.014~mag),
respectively. The amplitude ratios ($R_{k1}=A_k/A_1$) follow the changes of the
amplitudes but there is no significant change in the epoch-independent phase
differences ($\phi_{k1}=\phi_{k}-k\phi_{1}$) in accordance with the finding that no phase
modulation in the light curve changes has been detected.

\begin{figure}[t]
   \centering
   \includegraphics[width=8.8cm]{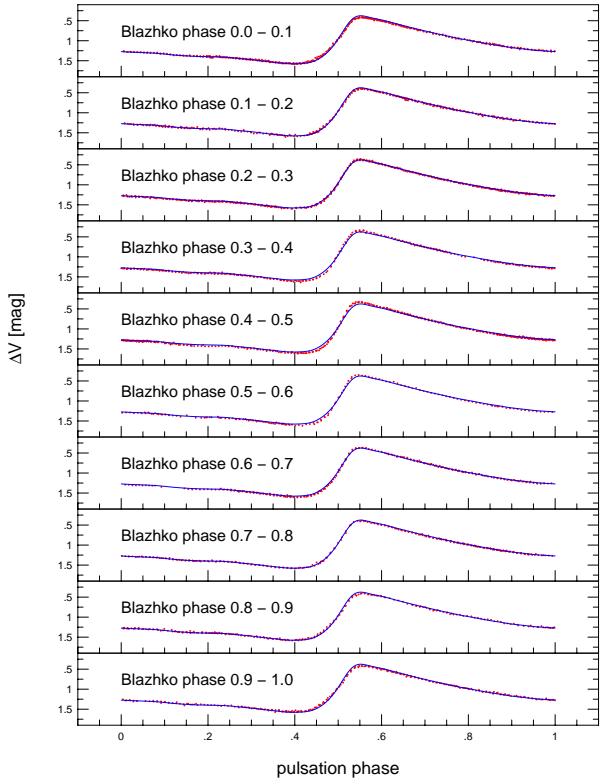}
\vskip 4mm
  \caption{Pulsation light curve constructed from data falling into the
0.1 phase intervals of the Blazhko cycle. In each plot the mean light curve
is also shown. Slight, but systematic differences of the observations in most of
the plots are evident.}
              \label{lcrrbl}
\end{figure}

\begin{figure}[t]
   \centering
   \includegraphics[width=8.8cm]{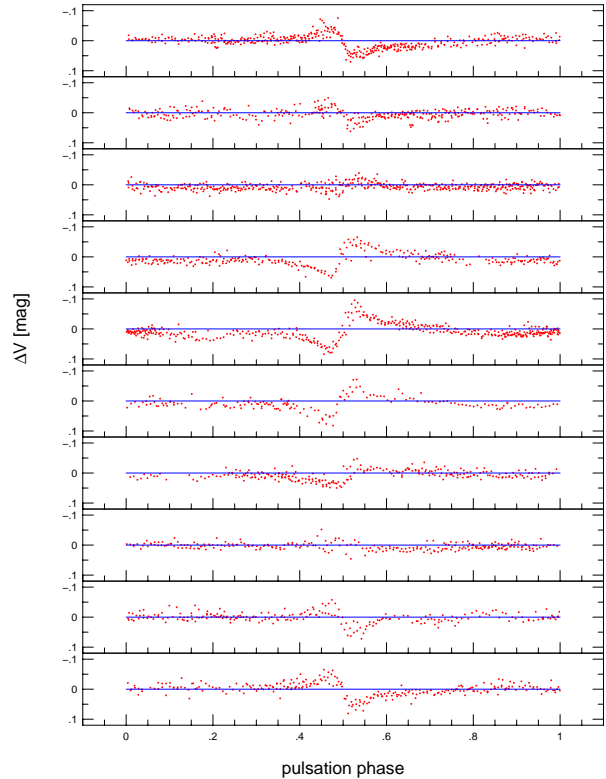}
\vskip 4mm
   \caption{The same data as shown in Fig.~\ref{lcrrbl} after subtracting the mean
light curve. Intense variations are concentrated in a 0.20 phase interval
centred on phase 0.5 which corresponds in our representation to the middle of the
ascending branch, when H emission is supposed to occur.}
              \label{lcrrres}
\end{figure}

We have checked how the light curves at different Blazhko phases
deviate from theo mean light curve. Fig.~\ref{lcrrbl} and \ref{lcrrres} 
show the pulsation light curves at each one-tenth phase of the Blazhko cycle,
and the residual curves after subtracting the mean light curve from the data
subsets.

Due to the good coverage of the data, a least squares fit with the pulsation period 
and its harmonics to the entire dataset (as shown in Fig.~\ref{rrgfold}) is a
good representation of the mean light curve of the variable. The Fourier
parameters of this `average' solution agree within the limits of the
uncertainties with the corresponding amplitudes and phases of the complete
solution listed in Table~\ref{four}. 


\begin{figure}[]
   \centering
   \includegraphics[width=8.8cm]{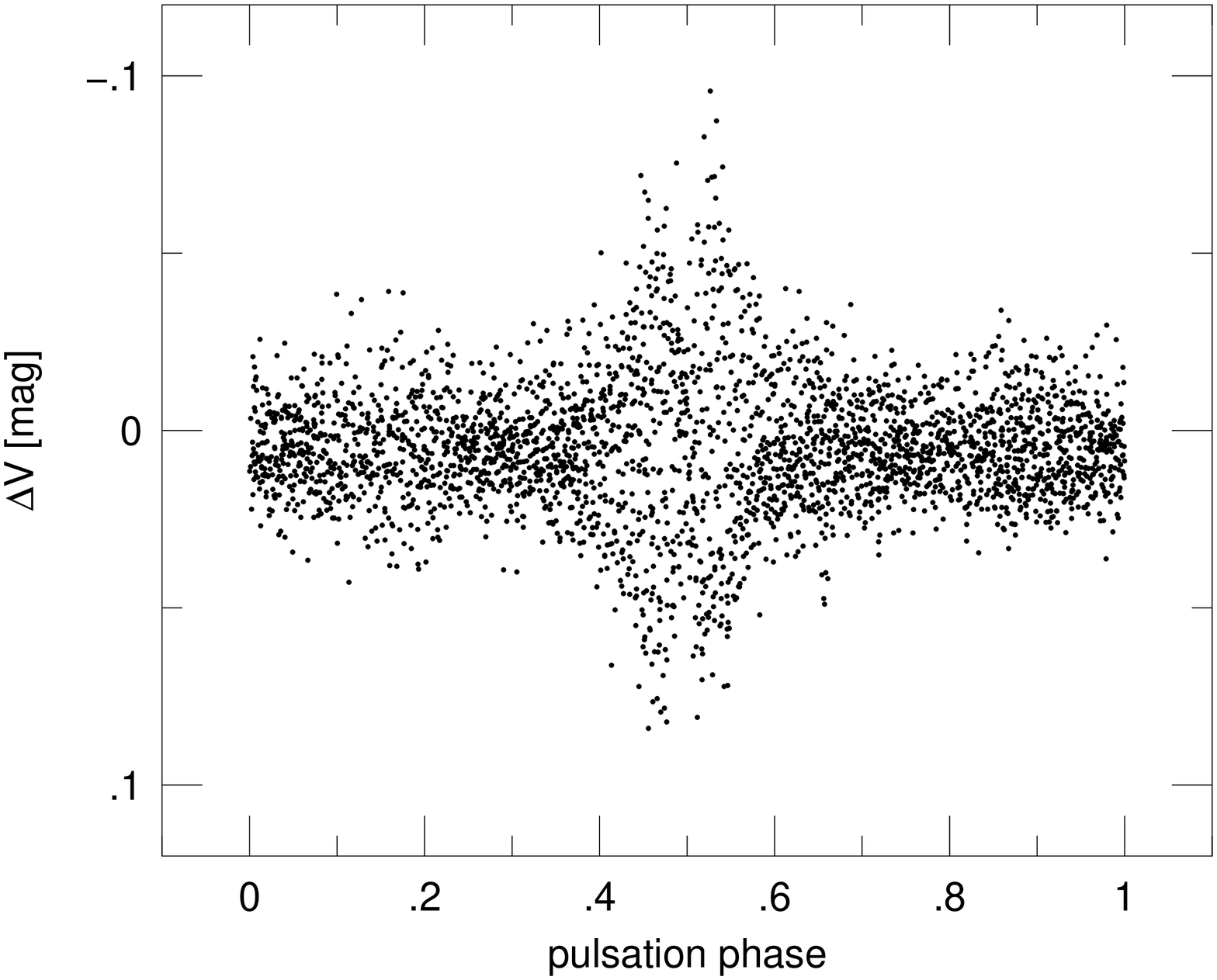}
   \caption{The $V$ residual light curve of \object{RR~Gem} folded with the
$0\fd3972884$ pulsation period after subtracting the mean pulsation light curve.
}
              \label{rrgresfold}
\end{figure}

The most striking property of the residual curves is that they show intense
variations only in a  narrow, $\sim$~$0.20$ phase interval of the pulsation. The
residual curve of all the data (Fig.~\ref{rrgresfold}) shows that the distortion
is symmetrical to phase 0.5 which is set to the phase of the middle of the rising
branch as described in Sect.~3. A similar residual plot has been shown by
\citet{leesm} for \object{DR~And} with the conclusion that `large residuals are
obtained from data points on the ascending branch'. To analyse such a behaviour
with the Fourier decomposition method might not be the most appropriate tool, as
for a reliable model very high harmonic modulation frequencies are needed. This is
reflected in the significantly different amplitude decrease behaviour of the 
modulation frequencies from the amplitude decrease of the harmonic components of the
main pulsation (see Fig.~\ref{amp2}). Although no complete fully nonlinear model
calculation exists, \citet{nowak2} showed that in evolved stars the nonradial
modes are highly nonlinear. However, even in the case of high nonlinearity it is hard
to explain why the high ($5-10$) order nonlinear coupling terms have amplitudes 
commensurable with the first order ones.

It is also important to think about why the modulation is concentrated in 
the minimum to maximum phase interval of the pulsation, where nonlinear effects
are most important. In the pioneering spectroscopic study of
\object{RR~Lyrae}, \citet{preston65} concluded that a displacement of the shock-forming 
layer in the atmosphere takes place during the Blazhko cycle. In
\citet{preston64} it was also shown that: `the phases at which H emission occurs
are closely related to phases of photometric parameters'. Thus the phase of the
$U-B$ excess and the  phase on the ascending branch when the visual luminosity of
the star  equals its time average value (at phase 0.5 in the figures)
are closely related, with only some minute differences, to the onset of the H
emission. The $U-B$ observations of \citet{liu89} indicate strong excess around
the phase of the mid ascending branch in \object{RR~Gem}. In Figs.~\ref{lcrrres} and
\ref{rrgresfold} the symmetrical modulation is centred exactly on this phase of
the pulsation, indicating a connection between the origin of the modulation and that
of the H emission. 

\begin{figure}[]
   \centering
   \includegraphics[width=8.8cm]{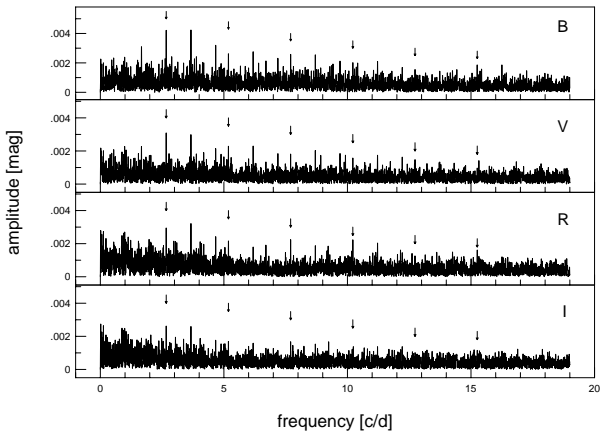}
   \caption{Amplitude spectra of the residual $B, V, R, I$ light curves
after prewhitening with the pulsation and modulation frequencies 
listed in Table~\ref{four}. Arrows indicate residual frequencies at 
$kf_0+f_{m}'$~($k=1..6$).
}
              \label{rrgsp1}
\end{figure}

As the standard deviations of the data from the fits using the 46 frequencies
listed in Table~\ref{four} are larger in each colour than expected from the
observations of the comparison and check stars, some further variations of the
residual light curves are also suspected. The Fourier spectra of the residuals
after prewhitening with both the pulsation and modulation frequencies listed in
Table~\ref{four} for the $B, V, R, I$ data are shown in Fig.~\ref{rrgsp1}.
Another modulation frequency component is present in each colour; the arrows
indicate the positions of $f=kf_0+f_m'$~$(k=1..6)$. This tertiary periodicity
corresponds to modulation with a period of $6\fd69$, which is very close to the
dominant modulation period. A tertiary period shorter than the dominant
modulation period has been also found by \citet{lacluyze} for \object{XZ~Cygni}.
A quintuplet structure with equidistant frequency spacing would predict secondary
modulation with a much shorter, $3\fd61$ periodicity.

\begin{table*}[tttthhhhhhhhh!!!]
   \caption{Magnitude and intensity mean brightnesses and colours in the different
Blazhko phases.}
         \label{col2}
\begin{tabular}{l@{\hspace{8pt}}r@{\hspace{8pt}}c@{\hspace{4pt}}c@{\hspace{4pt}}
c@{\hspace{4pt}}c@{\hspace{10pt}}c@{\hspace{5pt}}c@{\hspace{5pt}}c@{\hspace{10pt}}
c@{\hspace{4pt}}c@{\hspace{4pt}}c@{\hspace{4pt}}c@{\hspace{10pt}}c@{\hspace{5pt}}
c@{\hspace{5pt}}c}
            \hline\hline
            \noalign{\smallskip}
\multicolumn{1}{c}{}&&\multicolumn{7}{c}{magnitude averages}&\multicolumn{7}{c}{intensity
averages}\\
            \noalign{\smallskip}
            \hline
            \noalign{\smallskip}
Bl phase & $A_1(V)$ &  $\overline{B}$  &  $\overline{V}$ &
$\overline{R}$ &
$\overline{I}$ &
$\overline{B}-\overline{V}$ & $\overline{V}-\overline{R}$ &
$\overline{V}-\overline{I}$ & 
            $\overline{B}$  &  $\overline{V}$ &  $\overline{R}$ & $\overline{I}$ &
$\overline{B}-\overline{V}$ & $\overline{V}-\overline{R}$ &
$\overline{V}-\overline{I}$ \\
            \noalign{\smallskip}
            \hline
            \noalign{\smallskip}

$0.0-0.1$ & 0.3898 & 11.819 & 11.414 & 11.150 & 10.903 & 0.405 & 0.264 & 0.511 & 11.713 
& 11.359 & 11.118 & 10.885 & 0.354 & 0.241 & 0.474 \\
$0.1-0.2$ & 0.3981 & 11.818 & 11.416 & 11.150 & 10.905 & 0.402 & 0.266 & 0.511 & 11.707 
& 11.358 & 11.116 & 10.885 & 0.349 & 0.242 & 0.473 \\
$0.2-0.3$ & 0.4036 & 11.824 & 11.419 & 11.153 & 10.908 & 0.405 & 0.266 & 0.511 & 11.706 
& 11.358 & 11.117 & 10.888 & 0.348 & 0.241 & 0.470 \\
$0.3-0.4$ & 0.4104 & 11.827 & 11.421 & 11.156 & 10.910 & 0.406 & 0.265 & 0.511 & 11.702 
& 11.356 & 11.117 & 10.888 & 0.346 & 0.239 & 0.468 \\
$0.4-0.5$ & 0.4159 & 11.829 & 11.422 & 11.154 & 10.911 & 0.407 & 0.268 & 0.511 & 11.701 
& 11.355 & 11.115 & 10.888 & 0.346 & 0.240 & 0.467 \\
$0.5-0.6$ & 0.4088 & 11.824 & 11.421 & 11.155 & 10.908 & 0.403 & 0.266 & 0.513 & 11.701 
& 11.356 & 11.117 & 10.885 & 0.345 & 0.239 & 0.471 \\
$0.6-0.7$ & 0.4081 & 11.824 & 11.420 & 11.154 & 10.908 & 0.404 & 0.266 & 0.512 & 11.704 
& 11.357 & 11.117 & 10.886 & 0.347 & 0.240 & 0.471 \\
$0.7-0.8$ & 0.4953 & 11.823 & 11.416 & 11.152 & 10.905 & 0.407 & 0.264 & 0.511 & 11.710 
& 11.358 & 11.118 & 10.886 & 0.352 & 0.240 & 0.472 \\
$0.8-0.9$ & 0.3947 & 11.820 & 11.414 & 11.150 & 10.903 & 0.406 & 0.264 & 0.511 & 11.712
& 11.357 & 11.116 & 10.883 & 0.355 & 0.241 & 0.474 \\
$0.9-1.0$ & 0.3900 & 11.816 & 11.413 & 11.146 & 10.901 & 0.403 & 0.267 & 0.512 & 11.710
& 11.357 & 11.114 & 10.882 & 0.353 & 0.243 & 0.475 \\

            \hline
         \end{tabular}
   \end{table*}

\begin{figure}[]
   \centering
   \includegraphics[width=8.8cm]{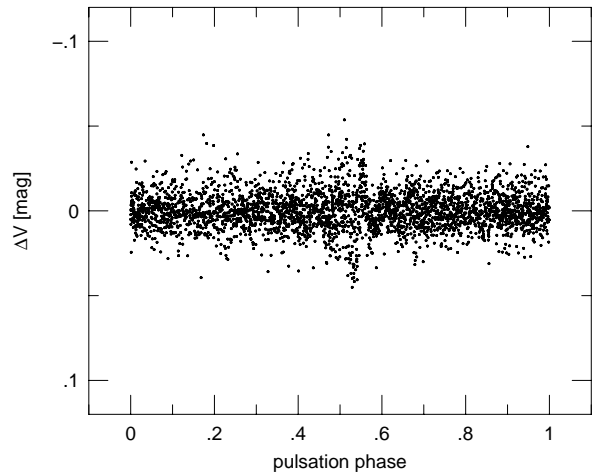}
   \caption{The $V$ residual light curve of \object{RR~Gem} folded with the 
$0\fd3972884$ pulsation period after prewhitening with the pulsation and
modulation frequencies listed in Table~\ref{four}.  Residual light variation is
still concentrated in the same ($0.4-0.6$) phase interval of the pulsation as in
Fig.~\ref{rrgresfold}.
}
              \label{rrgresf46}
\end{figure}

Fig.~\ref{rrgresf46} shows the residual $V$ light curve after prewhitening with
the frequencies of Table~\ref{four} folded with the pulsation period.
Residuals are still concentrated in the same phase of the pulsation 
as in Fig.~\ref{rrgresfold}. This behaviour of the residuals indicates 
that, perhaps, some irregular behaviour of the Blazhko modulation
mimics the tertiary periodicity.
This idea seems to be supported by separate analyses of the two halves of the
data, as neither parts of the data show the same residual structure shown in 
Fig.~\ref{rrgresf46}.

\subsection{Magnitude and colour changes}  

Multicolour photometry of Blazhko stars which were extended enough to determine
whether there is any change in the mean colour of the variables during the
Blazhko period has never been obtained.

According the mean characteristics of the Blazhko variables 
in the MACHO database, the modulation frequency occurs in the
spectra with an overall amplitude of 0.006~mag \citep{macho}, i.e., this 
is the typical amplitude of the the modulation in the magnitude average 
$V$ light level of Blazhko variables.

\begin{figure}[h]
   \centering
   \includegraphics[width=8.8cm]{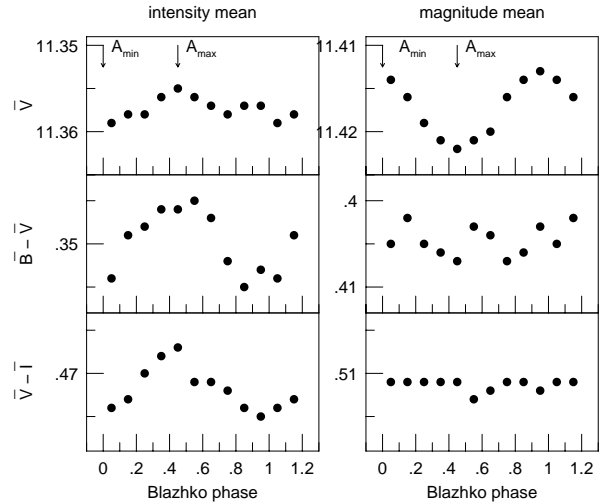}
   \caption{Intensity and magnitude mean $\overline{V}$ magnitude and
$\overline{B}-\overline{V}$ and $\overline{V}-\overline{I}$ colours of 
\object{RR~Gem} in the different Blazhko phases. The intensity mean colours
indicate  colour variations during the Blazhko cycle, namely the star is slightly
cooler in the smaller amplitude phase than during the highest amplitude period. 
}
              \label{amp}
\end{figure}

In Table~\ref{col2} the magnitude and intensity mean $B, V, R, I$ magnitudes
of \object{RR~Gem} in the different Blazhko phases are listed. Mean colours are 
calculated as the arithmetic differences between the mean magnitudes. 
Fig.~\ref{amp} shows the mean $\overline{V}$ magnitude and  
$\overline{B}-\overline{V}$ and $\overline{V}-\overline{I}$ colour
variations during the Blazhko cycle calculated as magnitude and
intensity averages. Although the magnitude mean brightness level of the star seems
to vary with a 0.005~mag semi-amplitude (in accordance with the MACHO results),
the intensity mean brightness shows much smaller if any variation. Concerning the
colours, the situation is the opposite; the colours calculated from the intensity
mean magnitudes definitely show correlated changes with the Blazhko phase,
namely, the star is slightly cooler in its small amplitude phase than
during the highest amplitude period.

Whether intensity or magnitude mean colors better correspond to the static values
of pulsating stars is still controversial \citep{bono}.
As the luminosities of the stars have real physical meaning,
we think that the intensity means reflect the real parameters of the star more
correctly. 
According to the intensity mean values the mean brightness level of
\object{RR~Gem} remains the same
during the Blazhko cycle, but a slight colour (temperature) variation
takes place in the sense that the star is $\sim30-40$K cooler when the
amplitude of the pulsation is the smallest.

\section{Discussion}

The extended CCD observations of \object{RR~Gem} exposed some previously unknown
or unexamined properties of the Blazhko phenomenon. The new results concerning
the properties of the modulation of \object{RR~Gem} are summarized in the next
items.

\begin{itemize}

\item
{First detection of a very low amplitude Blazhko modulation
with symmetrically placed triplets in the Fourier
spectrum.}

\item
{The modulation is concentrated in a 0.2 phase interval
of the pulsation, centred on the mid of the ascending
branch.}

\item
{In contrast to the exponential decrease in the
amplitudes of the harmonic components of the 
radial pulsation, the decrease in the amplitudes of
modulating components is close to linear.}

\item 
{In contrast to theoretical expectations the
amplitude ratios and phase differences of the
$\pm$ modulation components are not the same for 
the different order modulation component pairs and in
the different colours.}

\item
{According to the intensity mean colours and magnitudes
in the different phases of the Blazhko cycle,
there is very small if any mean $V$ brightness variation
along the Blazhko period, while the mean colours
indicate small but definite changes in the sense that
the star is 30-40~K cooler in its small amplitude
phase.}

\item 
{The light curve cannot be completely fitted with
symmetrically placed Fourier frequency components
alone. Residual scatter of the light curve is still
concentrated in the ascending branch,
indicating some irregular behaviour of the modulation.}

\end{itemize}    

Any model which relates the Blazhko periodicity
to the rotation of the star predicts a similar 
percentage of small amplitude modulation as
large amplitude ones. 
Because of the incidental angle of view of the rotational axes,
the observed amplitudes of the modulation has to vary uniformly 
from zero up to some tenths of a magnitude. On the contrary, 
no small amplitude Blazhko modulation has been known previously. 
This fact has to be carefully examined on the grounds of observational biases.
The most extended observations of field RR~Lyrae stars 
were the photographic observations from the middle of the $20^{th}$~century.  
These observations were not, however, accurate enough to detect modulation
amplitudes smaller than $\sim$~$0.2$~mag. Photoelectric observations of field RR~Lyrae
stars are very sparse and usually time limited, so again, they were not suitable to
detect small amplitude, long period modulations. The cluster and extragalactic 
variable surveys are usually seriously affected by crowding problems and are also 
often very time limited, thus small amplitude modulations might be 
missed or attributed to observational inaccuracy.

The detection of hundredths of magnitude amplitude modulation of \object{RR~Gem}
indicates that there might be a serious selection effect against small amplitude
Blazhko stars. A systematic search using today's accurate photometric detectors
is needed to reveal the true unbiased statistics of the modulation amplitudes.

Among the 731 fundamental mode Blazhko variables of the MACHO database
there was only one star with a modulation period shorter than 10 days
and only 5 with $P_{Bl}<15$~days \citep{macho}. In contrast, \object{RR~Gem} is
already the second very short modulation period variable in the much smaller
sample of well observed galactic field Blazhko variables (less than 50 stars).
The other short period Blazhko star (\object{AH~Cam}) was studied in detail by
\citet{smithah}. Both \object{AH~Cam} and \object{RR~Gem} are relatively metal
rich variables with pulsation periods shorter than $0\fd4$. These are their common
features, but the modulation pattern of the two stars are very different. 
In \object{AH~Cam} the amplitude of the modulation is $0.3-0.4$~mag, and
phase modulation with about 0.1 pulsation phases occurs as well. 
A simplified explanation of the larger percentage of short modulation
period Blazhko stars in our Galaxy than in the LMC would relate the
short modulation periods to the larger metal content of these stars. 
However, the 530 day modulation period of \object{RS~Boo} (\cite{kanyo, nagy})
with $P_{puls}=0.377$~d and [Fe/H]=$-0.35$~dex \citep{layden} indicates that a
more complex description is needed.

Though \citet{chadid}, favouring the nonradial mode explanation of the Blazhko
phenomenon, argue against the presence of a strong magnetic field 
in \object{RR~Lyrae}, our detailed analysis of \object{RR~Gem} revealed a challenge
to these models as well. The two most important results which would be hard to
explain by nonradial modes are the followings.

a) The linear decrease of the amplitudes of the modulation components  
compared to the exponential decrease in the amplitudes of the 
radial mode (Fig.~\ref{amp2}) indicates significant differences
between these frequency components. As nonlinearity is
proportional to the amplitude, it is expected that 
if the harmonic terms of the radial mode invoked by nonlinear 
coupling decrease exponentially, then the nonlinear coupling terms 
of the much smaller amplitude nonradial modes
have to decrease even more steeply.
The linear decrease of the amplitudes of the very small amplitude modulation
components indicates a different origin of its higher order components than
nonlinear coupling. This anomalous behavior of the amplitudes of the modulation
frequencies  arises from the high concentration of the modulation in
only 0.2 phase intervals of the pulsation. 

b) The modulation is centred on the phase when H emission is assumed to occur.
Such a connection between a temporal physical phenomenon in the atmosphere
and the occurrence of the modulation suggests that 
during the Blazhko period real changes in the atmosphere are present.
This is also supported by the detected colour variations during the Blazhko cycle.

Our detailed photometric observations of \object{RR~Gem} have given new insight into
the Blazhko phenomenon. However, it is an open question whether the Blazhko
properties of \object{RR~Gem} are unique or give general information about the
modulation as well. To find the answer, further similar studies of other Blazhko
variables are planned.

\begin{acknowledgements}
This research has made use of the SIMBAD database, operated at CDS Strasbourg,
France. The financial support of OTKA grants T-043504 and T-046207 is
acknowledged.

\end{acknowledgements}

\bibliographystyle{aa}

\end{document}